\documentclass[12pt,preprint]{aastex}

\usepackage{hyperref}
\usepackage{graphicx}
\usepackage{natbib}
\usepackage{amsmath,amsthm,amssymb}
\usepackage{lscape}

\def\spose#1{\hbox to 0pt{#1\hss}}

\def\kms{\ifmmode {\rm\,km\,s^{-1}}\else
    ${\rm\,km\,s^{-1}}$\fi}
\def\kmsMpc{\ifmmode {\rm\,km\,s^{-1}\,Mpc^{-1}}\else
    ${\rm\,km\,s^{-1}\,Mpc^{-1}}$\fi}

\def\msun{\ifmmode {\rm\,M_\odot}\else ${\rm\,M_\odot}$\fi}
\def\Msun{\ifmmode {\rm\,M_\odot}\else ${\rm\,M_\odot}$\fi}
\def\lsun{\ifmmode {\rm\,L_\odot}\else ${\rm\,L_\odot}$\fi}
\def\Lsun{\ifmmode {\rm\,L_\odot}\else ${\rm\,L_\odot}$\fi}
\def\rsun{\ifmmode {\rm\,R_\odot}\else ${\rm\,R_\odot}$\fi}
\def\Rsun{\ifmmode {\rm\,R_\odot}\else ${\rm\,R_\odot}$\fi}

\def\cm{{\rm\,cm}}
\def\cm3{\ifmmode {\rm\,cm^{-3}}\else ${\rm\,cm^{-3}}$\fi}

\def\ergps{\ifmmode {\rm\,erg\,s^{-1}}\else ${\rm\,erg\,s^{-1}}$\fi}
\def\ergpscm2{\ifmmode {\rm\,erg\,s^{-1}\,cm^{-2}}\else
    ${\rm\,erg\,s^{-1}\,cm^{-2}}$\fi}

\def\microJy{\,\mu {\rm Jy}}

\def\deg{\ifmmode {^{\circ}}\else {$^\circ$}\fi}
\def\degr{\ifmmode {^{\circ}}\else {$^\circ$}\fi}
\def\degs{\ifmmode {^{\circ}}\else {$^\circ$}\fi}

\def\micron{\,\mu {\rm m}}

\def\h3Mpc{h^{-3}{\rm Mpc}^3}
\def\Ho{\ifmmode {\rm\,H_\circ}\else ${\rm\,H_\circ}$\fi}
\def\hnot{\ifmmode {\rm\,H_\circ}\else ${\rm\,H_\circ}$\fi}
\def\h0{\ifmmode {\rm\,H_\circ}\else ${\rm\,H_\circ}$\fi}
\def\hnotunit{\ifmmode {\rm\,km\,s^{-1}\,Mpc^{-1}}\else
    ${\rm\,km\,s^{-1}\,Mpc^{-1}}$\fi}
\def\qnot{\ifmmode {\rm\,q_\circ}\else ${\rm q_\circ}$\fi}
\def\q0{\ifmmode {\rm\,q_\circ}\else ${\rm q_\circ}$\fi}

\def\arcsec{\ifmmode {^{\prime\prime}}\else $^{\prime\prime}$\fi}
\def\asec{\ifmmode {^{\prime\prime}}\else $^{\prime\prime}$\fi}
\def\arcmin{\ifmmode {^{\prime}}\else $^{\prime}$\fi}
\def\amin{\ifmmode {^{\prime}}\else $^{\prime}$\fi}

\def\secper{\ifmmode \rlap.{^{s}}\else $\rlap{.}{^{s}} $\fi}
\def\minper{\ifmmode \rlap.{^{m}}\else $\rlap{.}{^m} $\fi}
\def\magper{\ifmmode \rlap.{^{m}}\else $\rlap{.}{^m} $\fi}
\def\arcsper{\ifmmode \rlap.{^{\prime\prime}}\else
    $\rlap.{^{\prime\prime}}$\fi}
\def\arcmper{\ifmmode \rlap.{^{\prime}}\else
    $\rlap.{^{\prime}}$\fi}
\def\spose#1{\hbox to 0pt{#1\hss}}
\def\simlt{\mathrel{\spose{\lower 3pt\hbox{$\mathchar"218$}}
     \raise 2.0pt\hbox{$\mathchar"13C$}}}
\def\simgt{\mathrel{\spose{\lower 3pt\hbox{$\mathchar"218$}}
     \raise 2.0pt\hbox{$\mathchar"13E$}}}

\def\apjl{{ApJ}}

\def\apjs{{ApJS}}

\def\apjref#1;#2;#3;#4 {\par\pp#1, {#2}, #3, #4 \par}

\shorttitle{High $b$ {\it WISE}}
\shortauthors{Lake et al.}

\begin{document}

\title{Optical Spectroscopic Survey of High Latitude {\it WISE} Selected Sources}
\author{S.~E.~Lake\altaffilmark{1}, E.~L.~Wright\altaffilmark{1}, S.~Petty\altaffilmark{1}, R.~J.~Assef\altaffilmark{2,3}, T.~H.~Jarrett\altaffilmark{4}, S.~A.~Stanford\altaffilmark{5,6}, D.~Stern\altaffilmark{2},  C.-W.~Tsai\altaffilmark{4}}

\altaffiltext{1}{Physics and Astronomy Department, University of California, Los Angeles, CA 90095-1547}
\altaffiltext{2}{Jet Propulsion Laboratory, California Institute of
Technology, 4800 Oak Grove Dr., Pasadena, CA 91109}
\altaffiltext{3}{NASA Postdoctoral Program Fellow}
\altaffiltext{4}{ Infrared Processing and Analysis Center (IPAC), California Institute of Technology, Pasadena, CA 91125, USA }
\altaffiltext{5}{Department of Physics, University of California, Davis, CA 95616}
\altaffiltext{6}{Institute of Geophysics and Planetary Physics, Lawrence Livermore National
Laboratory, Livermore CA 94551}

\email{lake@physics.ucla.edu}

\begin{abstract}
We report on the results of an optical spectroscopic survey at high Galactic latitude ($|b|\ge 30\degs$) of a sample of {\it WISE}-selected targets, grouped by {\it WISE} W1 ($\lambda_{\rm eff} = 3.4\micron$) flux, which we use to characterize the sources {\it WISE} detected.
We observed 762 targets in 10 disjoint fields centered on ultra-luminous infrared galaxy (ULIRG) candidates using the DEIMOS spectrograph on Keck II. We find $0.30\pm 0.02 $ galaxies $\mathrm{arcmin}^{-2}$ with a median redshift of $z=0.33\pm0.01$ for the sample with $\mathrm{W}1 \ge 120 \microJy$. The foreground stellar densities in our survey range from $0.23 \pm 0.07\ \mathrm{arcmin}^{-2}$ to $1.1 \pm 0.1\ \mathrm{arcmin}^{-2}$ for the same sample. We obtained spectra that produced science grade redshifts for $\ge 90\%$ of our targets for sources with W1 flux $\geq 120\microJy$ that also had $i$-band flux $\gtrsim 18 \microJy$. We used for targeting very preliminary data reductions available to the team in August of 2010. Our results therefore present a conservative estimate of what is possible to achieve using {\it WISE'}s Preliminary Data Release for the study of field galaxies.
\end{abstract}



\section{Introduction}
The {\it Wide-field Infrared Survey Explorer} ({\it WISE}) is an all-sky mid-infrared survey satellite that NASA launched on December 14, 2009. Operating simultaneously in four bands, centered at 3.4, 4.6, 12 and 22 $\micron$ (W1-W4, hereafter), {\it WISE} completed its first full coverage of the sky in late July 2010, its cryogenic mission ended in early October 2010, and the satellite was put into safe mode in February 2011. {\it WISE} will provide an IR atlas of the full sky hundreds of times deeper than IRAS containing hundreds of millions of targets (5$\sigma$ point source sensitivity is equal to or better than 0.08, 0.11, 1 and 6 mJy in the four passbands). A Preliminary Data Release covering 57\% of the sky took place on  April 14, 2011.  Prior to this release, the {\it WISE} team undertook several programs to characterize the survey, including a comparison of {\it WISE} and {\it Spitzer} sources at the ecliptic poles by \cite{Jarrett:2011} and a study of the extragalactic source counts in the Bo\"{o}tes field by \cite{Benford:2011}. 

We expect that {\it WISE} has detected cluster $L_\star$ galaxies out to a redshift of $z\sim1$. This expectation is based on using a passive evolution model from \cite{BC:2003} that fits cluster luminosity functions to predict the $L_\star$ flux density in W1 as a function of redshift. This puts {\it WISE} in an interesting position with respect to the publicly available spectroscopic surveys. The surveys with coverage comparable to {\it WISE} are not nearly as deep, and the deep surveys were not nearly as wide, as shown in Table \ref{T:surveys}. Furthermore, the targeting for most other surveys rely on mixes of limiting magnitudes in primarily optical bands, morphologies, and colors, increasing the problem of selection effects in any attempt to use them to characterize the {\it WISE} sources. While using existing spectroscopic databases is useful for determining the redshifts of numerous {\it WISE} sources, it is not possible to construct flux-limited samples in the {\it WISE} bands with complete spectroscopic coverage. The importance of having such data available is that it greatly simplifies the statistical analysis of quantities that rely on flux-limited of galaxy samples, such as luminosity functions or correlation functions. In order to characterize the sources selected by the W1 bandpass, we have carried out a spectroscopic survey of {\it WISE}-selected objects blind to all considerations but $\mathrm{W}1$ flux. Here we present the design and results from the survey we carried out using the DEep Imaging Multi-Object Spectrograph \citep[][DEIMOS]{DEIMOS} on the Keck II telescope on UT 2010 September 14 selected by a W1 flux-classified catalog.

\begin{deluxetable}{lccl}
	\tablewidth{0.72\textwidth}
	\tablecaption{Spectroscopic Survey Characteristics}
	\tablehead{ \colhead{Survey} & \colhead{$\mathrm{median}(z)$}  & \colhead{Coverage $(\Omega)$} &  \colhead{Ref } \\
	  &  & \colhead{$(\mathrm{sr})$}& }
	\startdata
		6dFGS & 0.053 & 5.2 & \cite{Jones:2009} \\
		SDSS-DR8 & 0.1 & 2.43 & \cite{SDSSdr8} \\
		2dFGRS & 0.11 & 0.5 & \cite{Colless:2003} \\
		WiggleZ & 0.6 & 0.3 & \cite{Drinkwater:2010} \\
		GAMA & 0.2 & $4.4\times 10^{-2}$ & \cite{Baldry:2010} \\
		AGES & 0.31 & $2.3\times 10^{-3}$ & \cite{AGES}\\
		DEEP2-DR3 & 0.76 & $1.1\times10^{-3}$ & \cite{Davis:2003}\\
		zCOSMOS & 0.61 & $5.2\times10^{-4}$ & \cite{Lilly:2007} \\
		This Work & 0.33 & $6.78\times10^{-6}$&
	\enddata
	\tablecomments{A sample of redshift surveys showing their area and depth. A more comprehensive comparison can be found in Figure 1 of \cite{Baldry:2010}. Note that while the areal coverage of our survey is small the 10 fields are widely dispersed, minimizing cosmic variance.}
	\label{T:surveys}
\end{deluxetable}

All magnitudes from the Sloan Digital Sky Survey (SDSS) are model fluxes converted to standard AB magnitudes, and all other magnitudes are given in the Johnson Vega system used in the {\it WISE} database. We made no attempt to adjust our photometry to account for source morphology or extent. Thus, colors based on combining {\it WISE} and SDSS magnitudes in this paper will have significant systematic offsets from the actual physical colors, but are useful nevertheless. We have also not corrected fluxes for extinction.

For this paper we assume a $\Lambda \mathrm{CDM}$ cosmology using the {\it WMAP} 7-year parameters found in \cite{WMAP7} with $\Omega_m = 0.266$,  $\Omega_\Lambda = 0.734$, and $H_0 = 71.0\ \rm{km}\ \rm{s}^{-1}\ \rm{Mpc}^{-1}$. All coordinates are listed in the J2000 reference frame.

\section{DEIMOS Survey Design}
To select our targets, we used the {\it WISE} Level 3 operations (L3o) preliminary database. The Infrared Processing and Analysis Center (IPAC) constructed the L3o database by coadding {\it WISE} frames in a stripe of ecliptic longitude from a single day. Each stripe had a depth near the center of approximately 12 frames of coverage; roughly equivalent to the depth of the Preliminary Data Release on the ecliptic. IPAC then extracted the source detections in the coadded images as described in section IV of the {\it WISE} Preliminary Release Explanatory Supplement, but with earlier versions of all software. The source extractions had a minimum signal to noise ratio (SNR) of 3.5 in one of {\it WISE'}s channels -- half that used for the Preliminary Data Release -- because reliability of the internal database was less of a concern than for the Preliminary Data Release.

The main selection criterion for objects in our survey was that the source was in the L3o database. We sorted these objects into three samples based on fluxes from {\it WISE}'s most sensitive band, W1: 1) the design required {\it WISE} sensitivity of $120\microJy$ \citep[5-$\sigma$, ][]{Liu:2008}, and 2) the initial estimate of the in-orbit sensitivity of $80\microJy$ \citep[5-$\sigma$, ][]{Wright:2010}. We will refer to these samples by the following names, inspired by their flux ranges in $\microJy$, from here on: $\{\ge 120\}$, $\{80$--$120\}$, and $\{<80\}$ (includes sources not detected in W1). The classification based on W1 flux determined the priority in resolving conflicts when assigning slits on the mask, maximizing our ability to construct complete W1 flux limited samples. We adopted the profile fit photometry (w1mpro) for our flux measurements because we expect the majority of our targets, stars and field galaxies, to be point-like in the {\it WISE} beam ($6.1\arcsec$ full width at half maximum [FWHM]). 

The only other selection criterion we imposed was that R $\geq 15.0$~mag, as suggested by the DEIMOS documentation, to avoid saturation of the detector. Out of a desire to maintain as wide a pool of targets as possible, we used the Naval Observatory Merged Astrometric Dataset (NOMAD)  to perform this cut. The imprecision of photographic magnitudes is acceptable here due to the rarity of such bright sources and the fact that they are already well characterized by extant surveys.

The survey fields all had near their center a {\it WISE}-selected high-$z$ ultra-luminous infrared galaxy (ULIRG) candidate from selection criteria addressed below. Since we expect that the rest of the field will be filled with objects at much lower redshift, this should not have significantly biased the survey. The presence of lower redshift contaminants, such as merging galaxies and AGN, in the color regions used to select these candidates does introduce a potential source of bias to our survey. We therefore include the ULIRG candidate selection criteria we used here even though we have found no evidence of such bias.

\begin{figure}[t]
	\begin{center}
	\includegraphics[width=\textwidth]{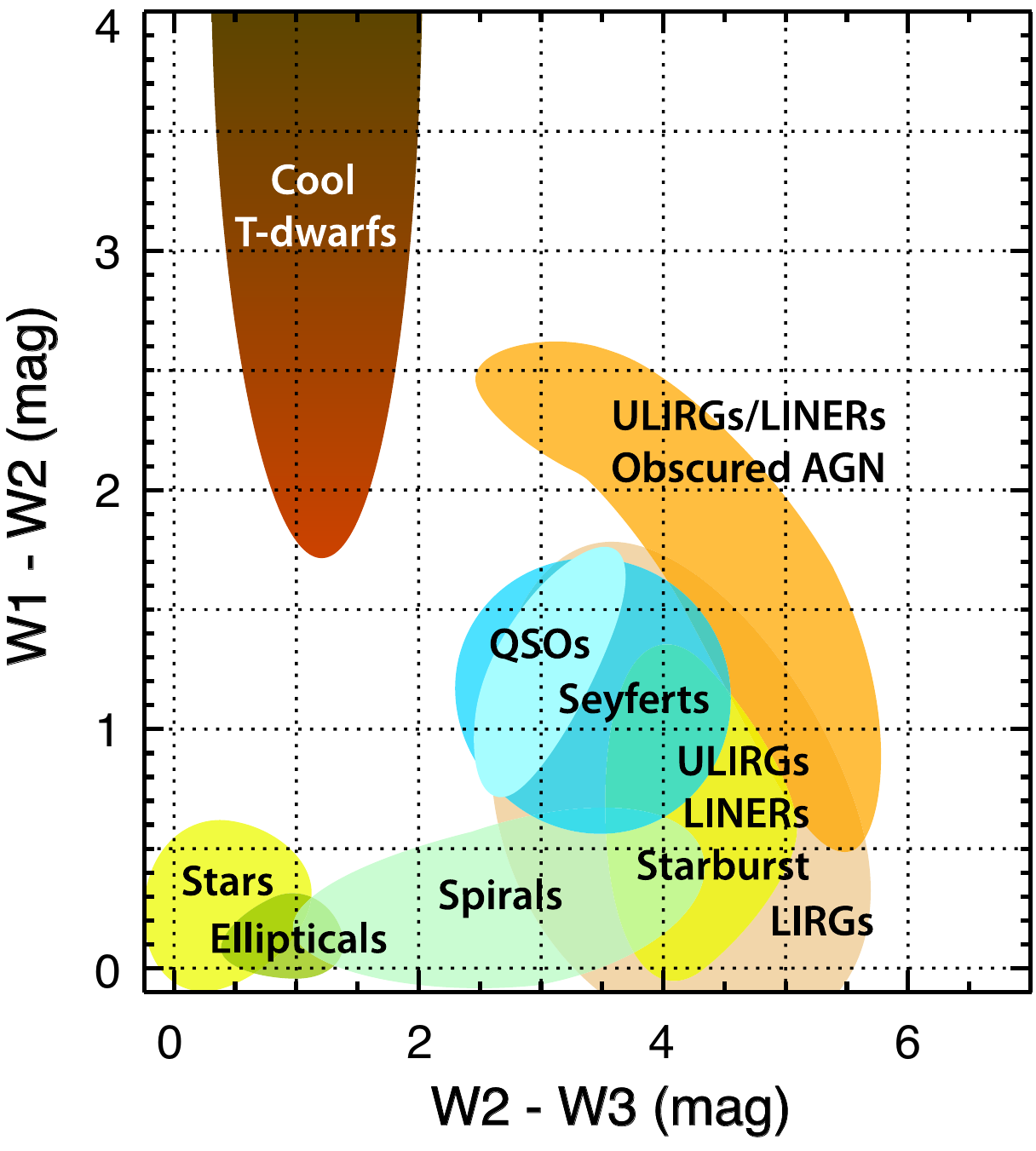}
	\end{center}
	
	\caption[{\it WISE} $\mathrm{W}1 - \mathrm{W}2$ versus $\mathrm{W}2 - \mathrm{W}3$ color space] { Color space regions where different classes of {\it WISE} detected sources, both Galactic and extragalactic, would fall based on SWIRE templates augmented by GRASIL models.}
	\label{F:Bubble}
\end{figure}

The {\it WISE} Extragalactic Team produced several different selection techniques for selecting ULIRGs and Hyper-Luminous Infrared Galaxies (HyLIRGS). One example of which can be found in the color-space plot in Figure \ref{F:Bubble}. In this figure we used templates from the Spitzer Wide-area InfraRed Extragalactic survey \citep[SWIRE][]{Polletta:2007} augmented by GRASIL models \citep{Silva:1998} to model the expected structure of the {\it WISE} $\mathrm{W}1 - \mathrm{W}2$ versus $\mathrm{W}2 - \mathrm{W}3$ color space. 

We selected ULIRG candidates based on the results of a previous spectroscopy study done using the Low Resolution Imaging Spectrometer \citep[LRIS, ][]{LRIS} at Keck I that followed up {\it WISE}-selected ULIRG candidates \citep[e.g.][]{Eisenhardt:2011}.
Of the 64 targets in the {\it WISE} team's LRIS observations as of August 2010, 18 were confirmed $z > 1.5$ 
ULIRGs. We used the {\it WISE} colors and magnitudes of these objects to
define a region of color-magnitude space to select the 10 ULIRG candidates for our
DEIMOS run. 
We used the {\it WISE} colors and magnitudes of the previously confirmed ULIRGs to produce the following limits: $0.3<\mathrm{W1}-\mathrm{W2}<2$, $\mathrm{W2}-\mathrm{W3} > 1.5$, $\mathrm{W3}-\mathrm{W4} > 2$, $\mathrm{W1}>10$, $\mathrm{W2}>10$, $\mathrm{W3}>8$, and $\mathrm{W4}> 5.5$~mag. The magnitude limits were imposed to bias the selection towards very red (i.e., $12$ and $22\micron$ bright) $z>1.5$ ULIRGs. The details of the LRIS observations are described in two upcoming papers: \cite{Eisenhardt:2011} and \cite{Bridge:2011}.

%

Each DEIMOS mask consists of a rectangular $16.7\arcmin \times 5\arcmin$ field from which two corners and a circular arc along the long side are lost to vignetting, leaving a total area of 68.3 $\mathrm{arcmin}^2$. Each target had a minimum slit length of $5\arcsec$ with $2\arcsec$ between the slits, allowing 70--80 targets per mask given the source densities in {\it WISE'}s L3o database. We designed the masks to have $2.0\arcsec$ wide slits in order to accommodate the astrometric uncertainties of the large number of fainter sources. As of September 2010, our understanding of the L3o database's astrometric accuracy could be found in \cite{Wright:2010}. They measured an astrometric uncertainty of $0.15\arcsec$ for {\it WISE} sources with $\rm{SNR}\ge20.0$. Extrapolating downward in quadrature as a means of estimation, this implies a $1$-$\sigma$ $1$-axis uncertainty of $0.62\arcsec$ for sources with $\rm{SNR}\sim 5$. With a $1\arcsec$ slit that implies a loss rate of $42\%$, whilst a $2 \arcsec$ slit gives $11\%$.

After positioning each field to contain the ULIRG candidate and 6 bright sources for alignment ($15 \le R \le 17$~mag), we assigned slits to objects in each of the categories using the dsimulator mask design software. Due to  the high sampling rate of targets in the $\{\ge120\}$ category we were able to include the good spectra for the alignment box targets without significantly biasing our results. All targets in a given W1 flux category had the same priority and thus the program assigned slits to them in such a way as to maximize the number of slits. We then added targets to the mask in order of decreasing category flux, resulting in the sampling rates given in Table \ref{T:samp}.  

\begin{deluxetable}{crrrrrcccc}
	\tabletypesize{\scriptsize}
	\rotate
	\tablewidth{0.95\textheight}
	\tablecaption{Field Characteristics}
	\tablehead{\colhead{Field} & \colhead{$\alpha$} & \colhead{$\delta$} & \colhead{PA} & \colhead{$b$} & \colhead{$\langle \mathrm{Coverage}\rangle$} & \colhead{$\{\ge 120\}$} & \colhead{$\{80$--$120\}$} & \colhead{$\{<80\}$} & \colhead{SDSS} \\
	\colhead{Number} & \colhead{$(\deg)$} & \colhead{$(\deg)$} & \colhead{$(\deg)$} & \colhead{$(\deg)$} & \colhead{(Frames)} & \colhead{$(f_{Q \ge 3}/f_{\mathrm{targ}}/N_{\mathrm{tot}})$} & \colhead{$(f_{Q \ge 3}/f_{\mathrm{targ}}/N_{\mathrm{tot}})$} & \colhead{$(f_{Q \ge 3}/f_{\mathrm{targ}}/N_{\mathrm{tot}})$} & \colhead{DR8}}
	\startdata
	$1$ & $310.28533$ & $-14.49725$ & $120$ & $-30.73209$ & 12.9 & $0.97/0.70/84$ & $1.00/0.29/34$ & $0.67/0.22/41$ & Y \\  
	$2$ & $312.36350$ & $-11.68394$ & $107$ & $-31.43735$ & 12.3 & $0.98/0.59/99$ & $0.90/0.30/33$ & $0.71/0.18/40$ & N \\  
	$3$ & $313.93537$ & $-12.49886$ & $73$ & $-33.17407$ & 11.5 & $0.98/0.77/74$ & $1.00/0.37/38$ & $0.83/0.29/21$ & P \\  
	$4$ & $338.96217$ & $16.07672$ & $86$ & $-35.70137$ & 12.8 & $0.96/0.72/68$ & $0.55/0.42/26$ & $0.81/0.27/59$ & Y \\  
	$5$ & $345.77133$ & $4.09244$ & $90$ & $-49.27066$ & 13.1 & $1.00/0.69/55$ & $0.89/0.36/25$ & $0.59/0.34/95$ & Y \\  
	$6$ & $25.06929$ & $-12.15469$ & $37$ & $-71.14944$ & 8.9 & $0.91/0.81/53$ & $0.85/0.38/34$ & $0.54/0.30/88$ & N \\  
	$7$ & $27.71867$ & $-18.08917$ & $161$ & $-73.59517$ & 7.0 & $0.97/0.77/44$ & $1.00/0.52/29$ & $0.60/0.24/103$ & P \\  
	$8$ & $38.37908$ & $23.55133$ & $-169$ & $-33.64597$ & 11.8 & $1.00/0.80/45$ & $0.91/0.46/24$ & $0.62/0.51/57$ & Y \\  
	$9$ & $47.65929$ & $12.24742$ & $-109$ & $-38.13031$ & 10.9 & $1.00/0.76/46$ & $0.70/0.51/39$ & $0.50/0.24/75$ & P \\  
	$10$ & $50.92692$ & $4.58761$ & $50$ & $-41.44942$ & 10.9 & $0.98/0.74/58$ & $0.72/0.49/37$ & $0.60/0.35/57$ & Y\\ \hline
	Combined &---&---& ---&---& $11.0$ &  $0.97/0.72/626$ & $0.84/0.41/319$ & $0.62/0.30/636$ & --- \\
	\enddata
	\tablecomments{The mask coordinates (J2000 equatorial coordinates), position angle, Galactic latitude, median coverage in $\mathrm{W1}$, whether the field overlaps with SDSS DR8, and fraction of slits that produced high quality spectra / fraction of targets assigned slits / total available targets, broken down by W1 flux sample (limits in $\microJy$). Per the recommendation made in the DEEP2 pipeline, the slits were set for $5\deg$ greater than the mask PA. The letters in the SDSS DR8 column denote whether the field overlaps with SDSS DR8, and stand for `Yes,' `No,' and `Partial.'}
	\label{T:samp}
\end{deluxetable}

We experienced a higher loss rate than anticipated due to the then-unknown pipeline error that led to quasi-random errors in declination in excess of the quoted positional uncertainty. Positions of objects in the {\it WISE} Preliminary Data Release may be offset from their true positions by many times the quoted positional uncertainty. Approximately 20\% of the sources fainter than $491\ \microJy$ suffer from a pipeline coding error that biases the reported position by $\sim0.2$--$1.0\arcsec$ in the declination direction. This error can affect sources as bright as $\mathrm{W1}\sim 2\ \mathrm{mJy}$. The effect of this error on slit losses can be mitigated by aligning slits away from an East-West PA. The Cautionary Notes section of the {\it WISE} Preliminary Release Explanatory Supplement describes the origin and nature of this effect in detail. 

In principle a $1\arcsec$ error combined with a $2\arcsec$ wide slit should not present a problem for sources brighter than $i\sim23$~mag with 45 minutes of integration on Keck. When combined with the astrometric uncertainty inherent in targeting sources with low SNR, though, it can result in the outright loss of a source. Some of our low SNR sources were as much as $4\arcsec$ away from the closest SDSS source. There are problems inherent in comparing surveys conducted at different wavelengths, so we did not perform a detailed analysis of all source offsets from SDSS counterparts. We will be able to better quantify how many sources were lost due to this problem after the final pass processing is complete. The aforementioned error will be corrected and the images will have greater coverage depth than the frames used to make the L3o database. Thus we will have a more uniform and accurate standard against which to compare the positions of targets used in this survey.

The primary focus of our survey was to obtain redshifts for the sources brighter than $80 \microJy$ in W1. We therefore decided to integrate for a total of 45 minutes on each field, splitting the time into 5-minute and 40-minute exposures to maximize our dynamic range. This strategy also allowed us to expose the maximum of 10 fields possible with DEIMOS in the course of a single night. The instrument documentation suggested that this exposure time would yield an SNR in the range of $3$--$7$ per pixel for a galaxy with $R_{\rm AB}= 21.0$~mag. We selected the lowest dispersion grating available in DEIMOS, $600\ \mathrm{lines}\ \mathrm{mm}^{-1}$, set for a $750.0$ nm central wavelength with the order blocking filter GG495. This produced a wavelength range of $495$--$1015$ nm for targets near the center line of the mask, and a moderate resolution. For objects filling the $2\arcsec$ slit we achieved a resolving power of $R= \lambda / \Delta \lambda = 861$ at $\lambda = 736.9\ \mathrm{nm}$. Exploring the resolving power across the wavelength range revealed that we were slit width limited, as shown by a strongly linear trend of the form $R = \lambda / \lambda_0$.

The artifact identification system was not used in the production of the L3o database we used for target selection, and we deliberately did not pre-reject sources based on visual identification on the chance that they may have been real sources that were contaminated. There was, therefore, definitely some contamination of the sample --- we later confirmed 9 targets as artifacts by examining the {\it WISE} images for targets that showed no optical emission and eliminated them from consideration in all metrics of the survey. To minimize the impact of this without introducing any non-flux cuts to the survey we positioned the fields to avoid very bright stars ($\mathrm{W1} \gtrsim 11.0$~mag) that produce large diffraction spikes.

We performed the data reduction using the DEEP2 pipeline, and the analysis using the SpecPro software package described in \cite{Masters:2011}. SpecPro calculates redshifts based on a cross correlation of templates to the 1-dimensional spectra, but we required clear identification of matching emission or absorption features to consider the redshift reliable. We calculated K-corrections using a weighted geometric mean of the low resolution spectral energy distribution (SED) templates of \cite{Assef:2010} as a crude approximation made for the sake of simplicity. Specifically, we used templates for elliptical, Sbc and Im galaxies and assigned them weights 0.5, 0.25 and 0.25, respectively. This approximation was made because nearly half of our sample did not have enough multi-wavelength broad-band photometry to perform proper SED fits to estimate K-corrections.

\section{Results and Discussion}
\begin{figure*}[t]
\begin{center}
	\includegraphics[width=\textwidth]{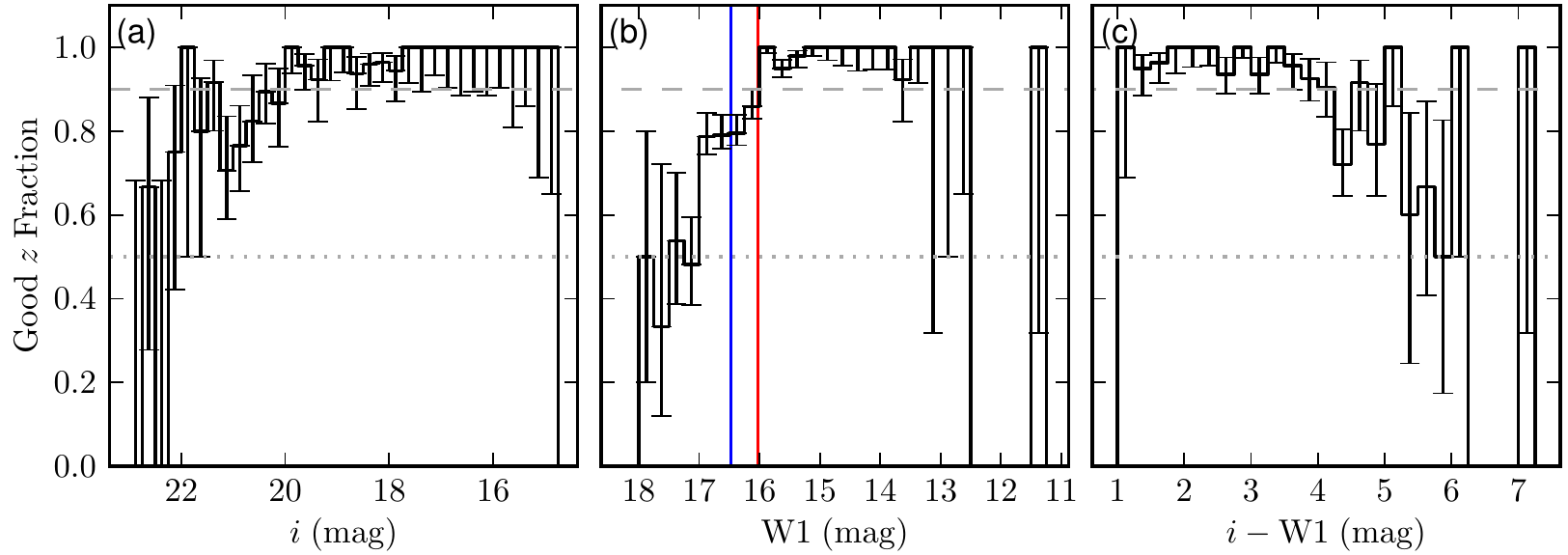}
	\caption{ The fraction of sources with magnitude uncertainty less than 1.0 in each magnitude bin that had $Q\ge3$ spectra. Panel a) shows the fraction of sources with good redshifts as a function of $i$ band AB model magnitudes using SDSS photometry. Panel b) shows the good redshift fraction as a function of $\mathrm{W}1$, with the sample boundaries highlighted using the solid vertical dark grey and light grey lines (colored blue and red for 80 and 120 $\microJy$, respectively, in the online version). Panel c) shows the same quantity as a function of $\mathrm{W1} - i$ color. Error bars assume that obtaining a good spectrum was a binomial random process, with confidence intervals constructed according to the technique in \cite{Feldman:1998}.  }
\label{F:depth}
\end{center}
\end{figure*}

We were able to successfully measure high quality redshifts based on visually identifiable emission or absorption features for 640 of the 762 targets in the 10 masks not lost to vignetting. Five of our fields fall within and three of them partially overlap with the Sloan Digital Sky Survey data release 8 \citep[][SDSS DR8]{SDSSdr8}, as seen in Table \ref{T:samp}. We therefore classified the fraction of targeted sources for which we obtained high quality spectra by flux density using both W1 and SDSS $i$ band model magnitudes in Figure \ref{F:depth}. This allows us to characterize the performance of our spectroscopic survey relative to the rest of the {\it WISE} survey for W1 and the sources contained in both {\it WISE} and SDSS for $i$.  Our survey successfully classified $>90\%$ of the target list for sources with $\mathrm{W1} \lesssim 16$~mag and $i \lesssim 20.75$~mag, and $>50\%$ for $\mathrm{W1} \lesssim 18$~mag and $i \lesssim 23$~mag. 

\begin{landscape}
\begin{figure*}[t]
\begin{center}
	\includegraphics[width=8.3 in]{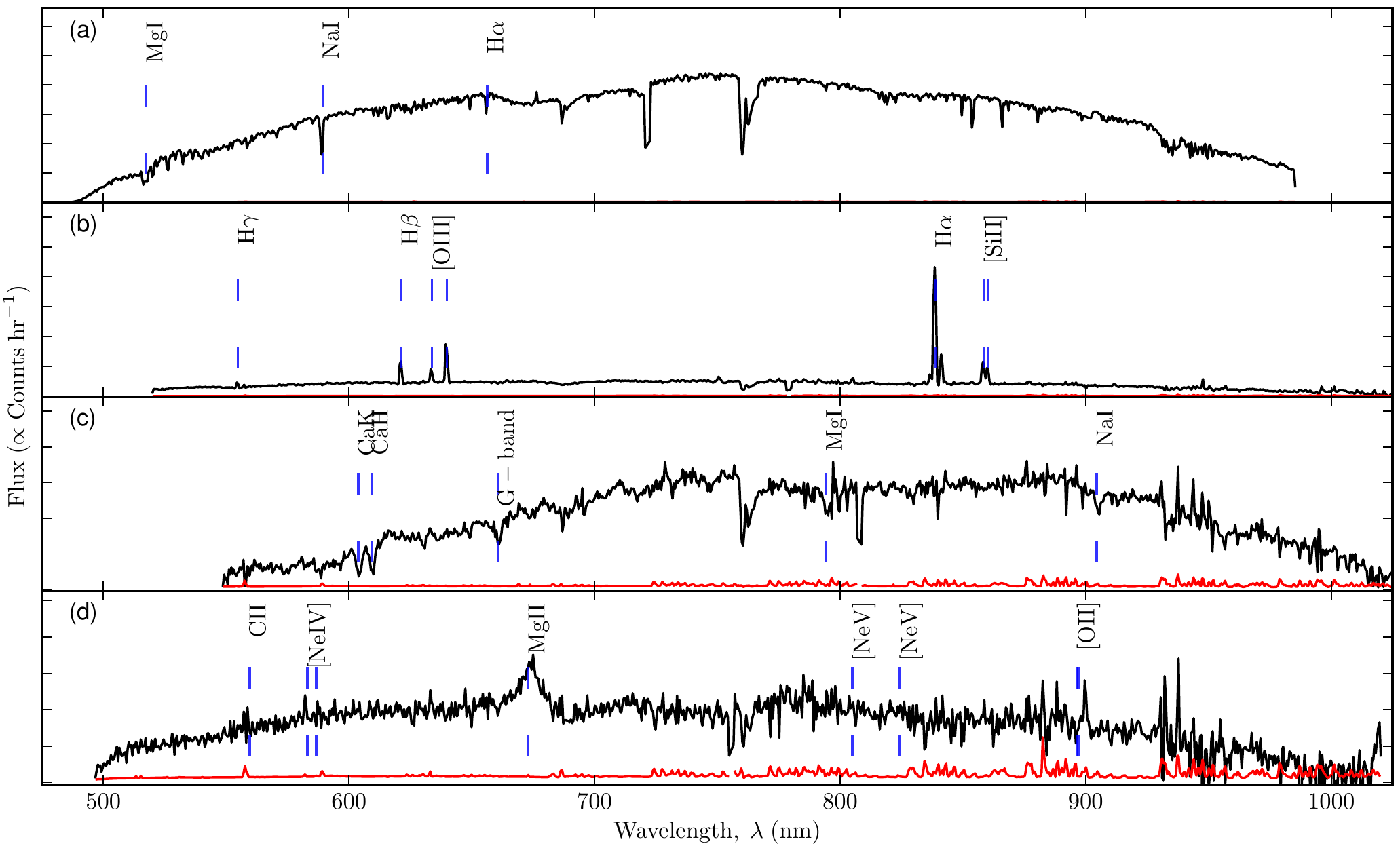}
	\caption{ Typical spectra we obtained for different classes of objects. All the 
	spectra are uncalibrated and binned down eight to one for clarity. All plots are in observed 
	frame wavelengths with typical spectral lines we used to identify such objects labeled with vertical 
	broken lines (colored blue in the online version), even if 
	the feature is not evident in the example spectrum. The line near the bottom of each graph is 
	the standard deviation of the flux at the given wavelength (colored red in the online version).
	Panel (a) is a spectrum for a K-type 
	star, (b) is an emission line galaxy at $z= 0.27802 \pm 0.00003$, (c) is an absorption 
	line galaxy at $z = 0.535 \pm 0.001 $, and (d) is a broad-lined AGN at 
	$z = 1.406 \pm 0.001$. All of the spectra shown are in the quality class $Q=4$.}
\label{F:egSpec}
\end{center}
\end{figure*}
\end{landscape}

In the following sub-sections we will discuss the properties of targets we were able to characterize using spectra such as those shown in Figure \ref{F:egSpec} and attempt to estimate the composition of those that we could not.  If the spectrum produced a science grade classification and redshift we assigned quality class, $Q$, of three or four. If the spectrum was inconclusive but confirmed the presence of a source then $Q$ is zero, one or two. If there was no evidence of emission in excess of the sky background anywhere, then $Q=-1$.

\subsection{Stellar Results}
\begin{figure}[t]
\begin{center}
	\includegraphics[width=\textwidth]{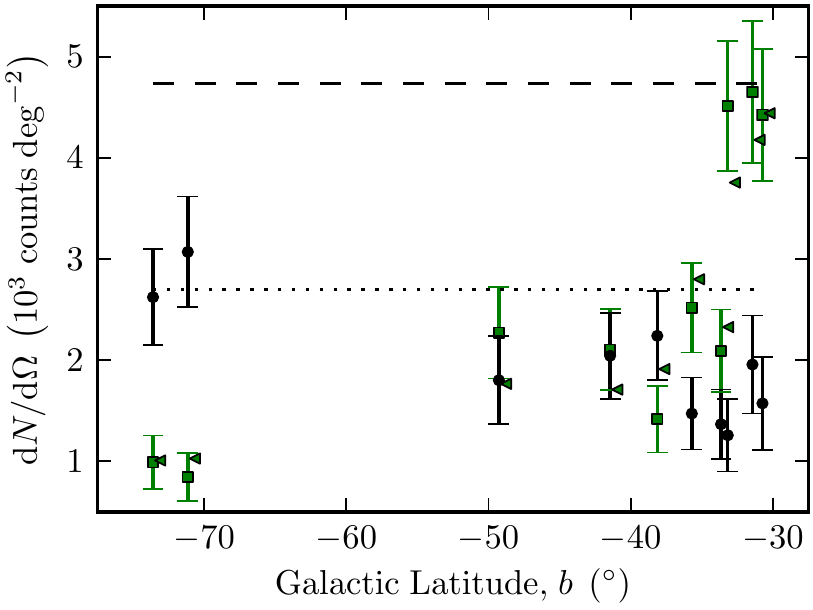}
	\caption{ Observed versus predicted source densities for sources with 
	W1 flux $\ge 80\microJy$. The black circles are the observed galaxy densities, while the 
	prediction is the horizontal dashed line and the 1-$\sigma$ uncertainty in that prediction 
	is the dotted line. The squares are the observed star densities, and 
	the triangles shifted to the right are the prediction (colored green in the online version). The 
	predicted star counts 
	are based on a model adapted from \cite{Jarrett:1994}. The predicted galaxy counts 
	come from integrating a Schechter luminosity function using the parameters measured in 
	\cite{Dai:2009} using the {\it Spitzer}/IRAC $3.6\micron$ channel out to a redshift of 1.25.}
\label{F:SGcts}
\end{center}
\end{figure}

\begin{figure}[t]
\begin{center}
	\includegraphics[width=\textwidth]{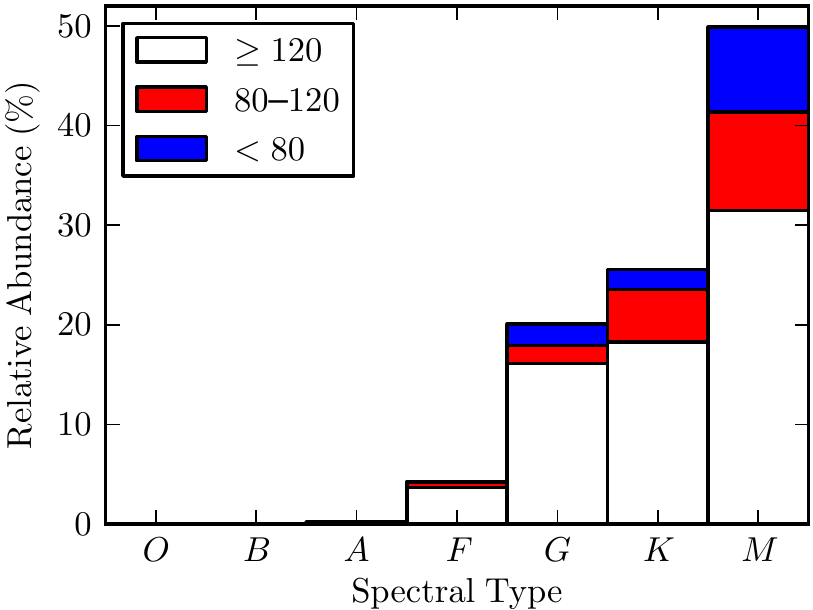}
	\caption{ Relative abundances of the different stellar spectral types. All quantities are attempt rate corrected. The accuracy of the spectral classification of 
	an individual star is accurate to at least $\pm$ half a spectral class. 
	The accuracy is better in the case of M-type stars 
	and stars that have good optical photometry from SDSS DR8. }
\label{F:SAbund}
\end{center}
\end{figure}

Figure \ref{F:SGcts} shows the stellar density for sources with W1 fluxes $\ge 80\microJy$ on a field by field basis, compared with the expected stellar contribution based on the Galactic star count model of \cite{Jarrett:1994}. The results are in good agreement with the model both overall and on a field by field basis. The breakdown of our target stars by spectral class is in Figure \ref{F:SAbund}. We also checked our stellar population for the occurrence of statistically significant color excesses in the longer wavelength channels of {\it WISE}. 
Of the 338 stars we spectrally identified, none of them presented a color $W1 - Wx \ge 1.0$ with the restriction that $\rm{SNR} \ge 7.0$ in both of the channels used to calculate the color. 

\subsection{Extragalactic Results}
\begin{figure}[t]
\begin{center}
	\includegraphics[width=\textwidth]{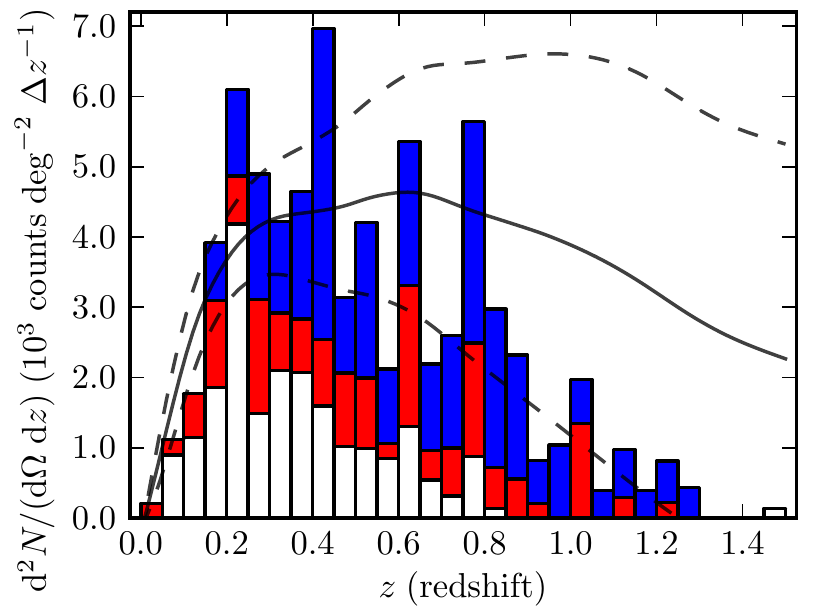}
	\caption{Stacked bar histogram of the robust spectroscopic redshifts from our survey. The 
		white bars are the $\{>120\}$ sample, the light grey (red in the online version) bars are the 
		$\{80$--$120\}$ sample, and the dark grey (blue in the online version) bars are the 
		$\{<80\}$ sample. The height of each 	
		bar was sampling rate corrected on a field by field basis by dividing by the values in  
		Table \ref{T:samp}. The solid line represents the predicted source density based on 
		integrating a Schechter function with the parameters from \cite{Dai:2009} down to an 
		observable flux limit of $80\microJy$. The dashed lines are the 1-$\sigma$ confidence 
		band for the prediction. }
\label{F:zhist}
\end{center}
\end{figure}

One simple measure of a photometric survey is the  median redshift of the galaxies detected. We estimate that {\it WISE} detects field galaxies back to the median redshift of at least $0.48\pm0.02$ by linearly interpolating the cumulative distribution that corresponds to Figure \ref{F:zhist}. The reason for referring to this value as a lower limit is due to the fact that we did not correct for the median lowering bias caused by reduced success rates in measuring redshifts of galaxies from the fainter samples. A more detailed breakdown of the median redshift by flux depth and galaxy spectral type can be found in Table \ref{T:exgal}. In principle, our results could also have been biased by the different spectral features present on the detector as a function of galaxy spectral type and redshift. For example, an absorption line galaxy with $z\lesssim 0.2$ would not have presented visible Ca II H and K lines, leaving us to rely on the less prominent Na I doublet at 580.3 nm and the Mg I line at 517.5 nm for measuring the redshift. The lesser prominence of these lines, however, means that we are more likely to be able to classify a faint source at higher redshift rather than at than lower, as evidenced by the absence of $\{<80\}$ sources in the three lowest bins of Figure \ref{F:zhist}.

\begin{figure}[t]
\begin{center}
	\includegraphics[width=\textwidth]{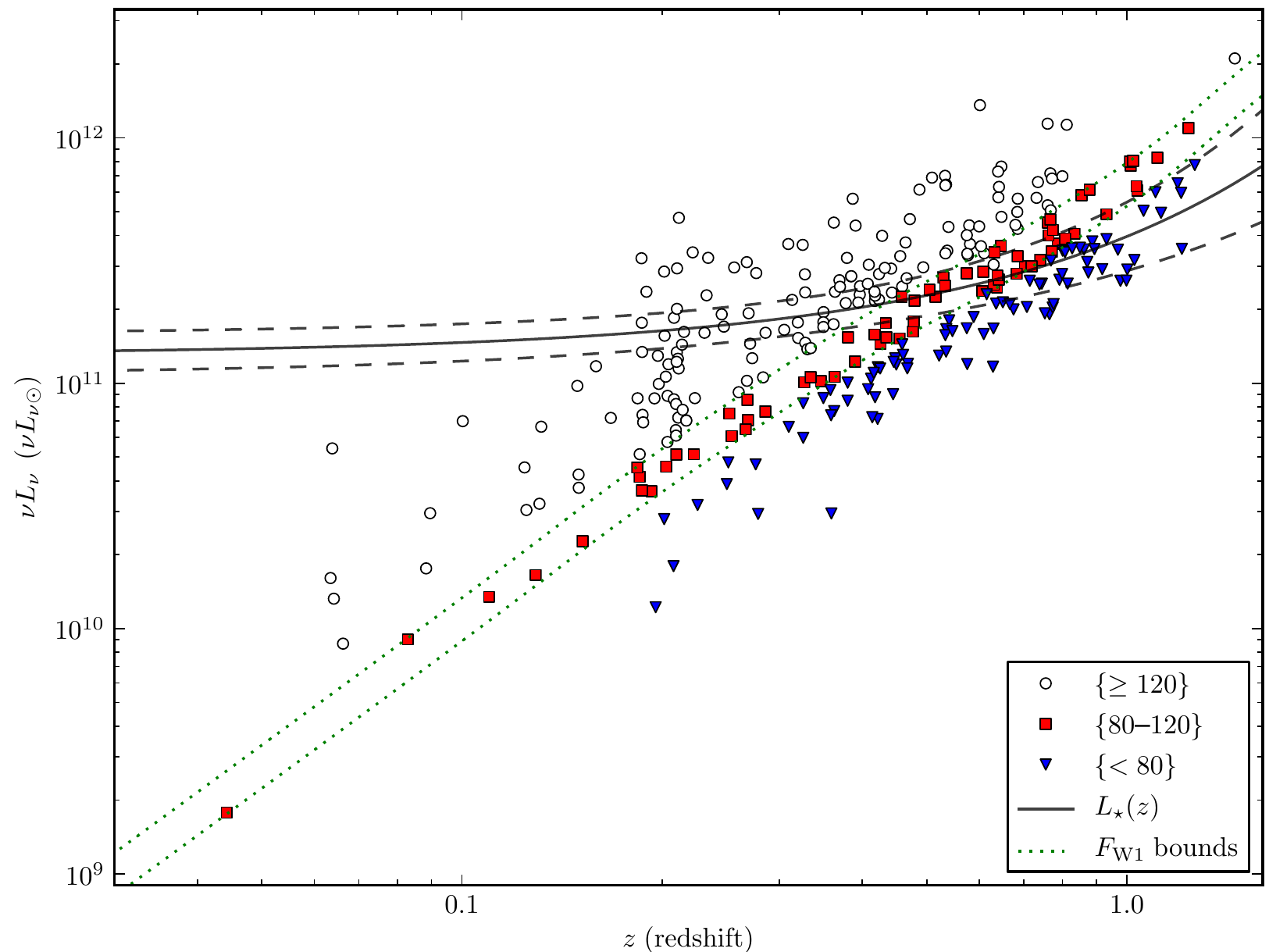}
	\caption{W1 luminosity, approximated as $\nu L_\nu$ in units where 
		$\nu L_\nu$ for Sol is 1, against observed 
		redshift. The solid line represents the expected $\nu L_\nu(z)$ for $L_\star$ galaxies 
		based on the linearly evolving $M_\star$ measurement from \cite{Dai:2009} using 
		the {\it Spitzer}/IRAC $3.6\micron$ channel. 
		The dashed lines are the $1$-$\sigma$ confidence bands from straightforward error 
		propagation of the uncertainties in the luminosity function parameters. The dotted lines
		are $\nu L_\nu(z)$ for an $120\microJy$ and $80\microJy$ source. The sources that
		cross into other regions are due to higher SNR photometry that became available 
		after target selection.}
\label{F:Lvz}
\end{center}
\end{figure}

A more important measure of the depth achieved, however, is the redshift to which $L_\star$ galaxies are detected in abundance, because this sets the depth to which the survey can provide constraints on the faint-end slope of the luminosity function. In Figure \ref{F:Lvz} we have plotted the luminosity of our sources versus their redshift alongside the simple evolving model for $L_\star(z)$ from \cite{Dai:2009} ($L_\star(z) = L_{\star0} 10^{0.4 Q z },\ Q=1.2\pm0.4$), making the approximation that the {\it Spitzer} InfraRed Array Camera (IRAC) channel 1, with effective wavelength $3.6 \micron$, is the same as W1. From this function for $L_\star(z)$ we find that at a W1 flux limit of $80\microJy$ {\it WISE} detects $L_\star$ galaxies back to $z=0.7^{+0.3}_{-0.2}$.

Our estimate of the redshift to which {\it WISE} detects $L_\star$ galaxies has a caveat to its accuracy. The evidence that a caveat is needed comes from the mismatch between the predicted galaxy densities and the observed densities, seen in Figures \ref{F:SGcts} and \ref{F:zhist}. In total {\it WISE} detects at least $(1.9 \pm 0.1) \times 10^3\ \mathrm{counts}\ \mathrm{deg}^{-2}$ field galaxies with observed $\mathrm{W}1$ flux $\ge 80 \microJy$ while the prediction from the model is $(5 \pm 2) \times 10^3\ \mathrm{counts}\ \mathrm{deg}^{-2}$, a discrepancy of $1.3\sigma$. Although the mismatch is not statistically significant, it is unlikely to be due to incompleteness in our survey, but possibly due to the extrapolation
of their results to a flux limit of $80\microJy$. The sample used by
\cite{Dai:2009} was limited to $[3.6] >143\microJy$ ($<15.7$ mag) and
their analysis was restricted to $z<0.6$. As discussed by \cite{Dai:2009}, 
the $M_\star$ evolution they find is faster than that found by
other surveys (although at different wavelengths) and seems to
overestimate that of other studies at higher redshifts (see their
Fig. 10). Coupled with their assumption of a non-evolving faint-end
slope (determined primarily from their $z<0.2$ sample), this could
tentatively explain the differences we find. A detailed study of
the WISE-selected mid-infrared galaxy luminosity function is the
subject of our follow-up publication, \cite{Lake:2011-2}.

Figures \ref{F:SGcts} and \ref{F:zhist}  also provide evidence that our ULIRG candidate targeting did not introduce a significant bias toward over-dense fields. Both the radial and angular densities show little evidence for the presence of large clusters. Indeed, the one significant single field over-density in redshift we found at $\langle z\rangle = 0.2127 \pm 0.0006$ was in field 6 ($b \approx -71\deg$) and the ULIRG candidate was at $z\approx 1$. 

\begin{figure}[t]
\begin{center}
	\includegraphics[width=\textwidth]{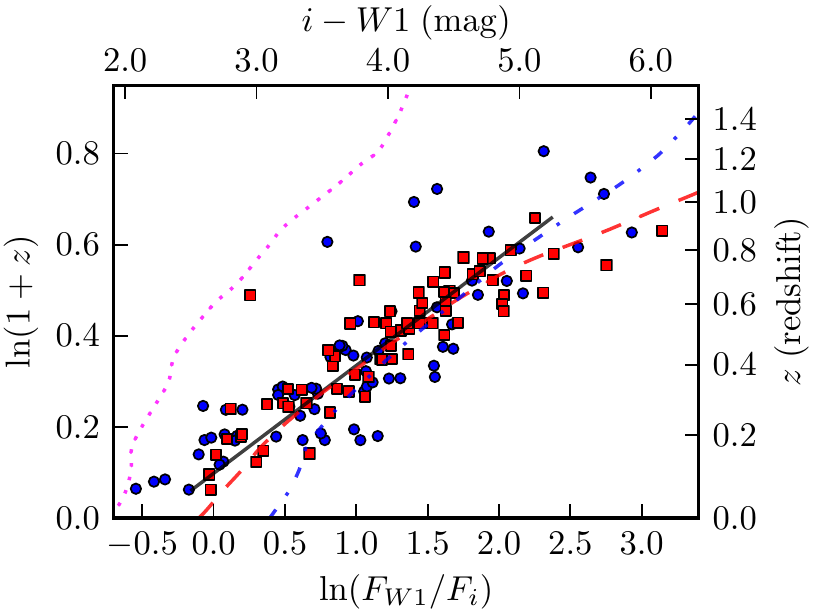}
	\caption{$\ln(1+z)$ versus $\ln(F_{\mathrm{W1}} / F_{i})$ (both fluxes in Jy) color for non-broad-lined 
	sources in our survey that have high quality ($Q>2$) spectra, unique source correlation between 
	SDSS and {\it WISE}, 
	and both $\sigma_i,\ \sigma_{\mathrm{W1}} \leq 1.0$~mag. The circles are galaxies for which we 
	detected emission lines (blue in the online version) and the squares had only absorption lines (red in the online version). The solid line is the result 
	of a maximum likelihood fit of color to a linear function of $\ln(1+z)$ with extrinsic scatter 
	for points with $z\le 1$. 
	The dashed, dash-dotted, and dotted lines are the tracks formed by the elliptical (red online), Sbc (blue online), 
	and Im (magenta online) templates from \cite{Assef:2010}, respectively.}
	\label{F:zcolor}
\end{center}
\end{figure}

We have found that the $1.6\micron$ bump in the typical galaxy SED produces a straightforward correlation between many of the {\it WISE}/SDSS colors and redshift for galaxies without observed broad-line emission. The correlation is strongest using the $g$, $r$, or $i$ filter with W1, but is apparent when matching any SDSS filter with W1 or W2. We plot an example in Figure \ref{F:zcolor}, using $i$ because it is the SDSS band nearest to the center of our spectra. We fit $\ln( F_{\mathrm{W}1} / F_i )$ to $\ln(1+z)$ using a function of the form $y = m (x - x_0) + b$ and a badness of fit/likelihood:
\begin{equation}
	-\ln(\mathcal{L}) = \frac{1}{2} 
		\sum_{i=1}^N \left[\frac{(y_i - m [x_i - x_0] - b)^2}{\sigma_{y,i}^2 + [ m \sigma_{x,i}]^2 + 
		\sigma_{\mathrm{ext}}^2 } + \frac{1}{2}\ln{\left(\sigma_{y,i}^2 + [ m \sigma_{x,i}]^2 + 
		\sigma_{\mathrm{ext}}^2 \right) } \right],
\end{equation}
where $\sigma_{\mathrm{ext}}$ is the extrinsic scatter in excess of the statistical uncertainty of the measurements and it, alongside $m$ and $b$, is a fit parameter on which we optimize $\mathcal{L}$. We set $x_0$ as a free parameter to reduce the off-diagonal terms of the covariance matrix of the main fit parameters. Specifically, we set $x_0 = \left\langle x_i \right\rangle$ with weights $w_i = \left(\sigma_{y,i}^2 + [ m \sigma_{x,i}]^2 + \sigma_{\mathrm{ext}}^2\right)^{-1}$.
The resulting parameters can be found in Table \ref{T:fit}. 
 The implication of this correlation is that Figure \ref{F:depth}c) gives us a rough estimate of our completeness relative to {\it WISE} as a function of redshift. Specifically, we are $90\%$ complete to $z\sim 0.5$ and $50\%$ complete to $z\sim0.9$. 

We present in Table \ref{T:zcat} an excerpt of the redshift catalog, made available as an electronic table with the online version. The table will only include science grade, $Q\ge 3$, redshifts derived from spectra.

\begin{deluxetable}{c l c c c}
	\tablewidth{0.90\textwidth}
	\tablecaption{Extragalactic Summary}
	\tablehead{ \colhead{Galaxy} & \colhead{W1 Flux } & \colhead{ Density } & 
				\colhead{$\mathrm{median}\ z$\tablenotemark{a}} & \colhead{Source} \\
		\colhead{Spectral Type} & \colhead{$(\microJy)$} & \colhead{$(\mathrm{count}\ \mathrm{arcmin}^{-2})$} & & \colhead{count} }
	\startdata
		Absorption\tablenotemark{b} & $\ge 120$ & $0.18 \pm 0.02$ & $0.36\pm0.01$ & 90 \\
			 & $\ge 80$ & $0.30 \pm 0.03$ & $0.41^{+0.03}_{-0.02}$ & 125 \\
			 & $\ge 0$ &  $0.39 \pm 0.04$ & $0.49^{+0.02}_{-0.04}$ & 143 \\
		Emission\tablenotemark{c} & $\ge 120$ & $0.12 \pm 0.02$ & $0.28\pm0.03$ & 60 \\
			 & $\ge 80$ & $0.24 \pm 0.03$ & $0.33^{+0.04}_{-0.02}$ & 94 \\
			 & $\ge 0$ & $0.60 \pm 0.05$ & $0.48\pm0.03$ & 165 \\
		All Field\tablenotemark{d}  & $\ge 120$ & $0.30 \pm 0.02$ & $0.33\pm0.01$ & 150 \\
			 & $\ge 80$ & $0.54 \pm 0.04$ & $0.39\pm0.02$ & 219 \\
			 & $\ge 0$ & $1.00 \pm 0.06$ & $0.48\pm0.02$ & 309 \\
		Broad-lined AGN & $\ge 120$ & $0.022 \pm 0.007$ & $0.5^{+0.3}_{-0.1}$ & 11 \\
			& $\ge 80$ & $0.028 \pm 0.008$ & $0.83^{+0.04}_{-0.3}$ & 13 \\
			& $\ge 0$ & $0.05\pm 0.01$ & $0.9^{+0.2}_{-0.1}$ & 17
	\enddata
	\tablenotetext{a}{Calculated by linearly interpolating the
		cumulative counts histogram using the same bins as in Figure \ref{F:zhist}.}
	\tablenotetext{b}{Galaxies without detected emission lines.}
	\tablenotetext{c}{Galaxies with detected emission lines that are not obviously broadened.}
	\tablenotetext{d}{Union of Absorption and Emission galaxies.}
	\tablecomments{Extragalactic population density and median redshift as a function of galaxy 
		type and W1 flux cutoff. The densities cannot be obtained by dividing the target count 
		column by the survey area because of the sampling rate corrections needed. }
	\label{T:exgal}
\end{deluxetable}

\begin{deluxetable}{clllc}
	\tablecaption{$i-\mathrm{W}1$ vs. $\ln(1 + z)$ Fit Parameters}
	\tablehead{ \colhead{$y$-variable} & \colhead{ slope ($m$) } & \colhead{ $y$-intercept ($b$) } & 
		\colhead{Scatter ($\sigma_{\mathrm{ext}}$)} & \colhead{ $x$-offset ($x_0$) } }
	\startdata
		$\ln(F_{\mathrm{W1}}/F_i)$ & $4.2\pm0.2$ & $1.05\pm0.03$ & $0.33\pm0.02$ & 0.364 \\
		$i - \mathrm{W}1$ & $4.6\pm0.2$~mag & $3.81\pm0.03$~mag & $0.36\pm0.02$~mag & 0.364 
	\enddata
	\label{T:fit}
\end{deluxetable}

\begin{deluxetable}{ccccccccc}
	\tabletypesize{\scriptsize}
	\rotate
	\tablewidth{0.86\textheight}
	\tablecaption{WISE-DEIMOS Redshift Catalog (Excerpt)}
	\tablehead{  \colhead{Designation} & \colhead{ $\alpha$ } & \colhead{ $\delta$ } & 
				 \colhead{$z$}& \colhead{$\sigma_z$} & \colhead{Quality} & 
				\colhead{SpecPro} & \colhead{$\mathrm{W}1$\tablenotemark{a}} &
				 \colhead{$\sigma_{\mathrm{W}1}$}   \\
			 & \colhead{ $(\deg)$ } & \colhead{ $(\deg)$ }  
				&  &  & \colhead{Class} & 
				\colhead{Template} & \colhead{(mag)} & \colhead{(mag)} }
	\startdata
		WISEPC J204116.72-143132.2 & 310.3196716 & -14.5256138 & 0.088342 & 0.000404 & 4 & VVDS Spiral & 15.458 & 0.054 \\
		WISEPC J204128.23-143052.4 & 310.3676453 & -14.5145798 & 0.402037 & 2.34e-05 & 4 & SDSS Quasar & 15.667 & 0.06 \\
		WISEPC J204115.66-143034.9 & 310.3152771 & -14.5097055 & 0.57926 & 0.000282 & 4 & VVDS Elliptical & 15.682 & 0.06 \\
		WISEPC J204125.02-143016.2 & 310.3542786 & -14.5045099 & 0.320373 & 0.000141 & 4 & VVDS Starburst & 15.833 & 0.067 \\
		WISEPC J204128.43-142940.3 & 310.3684998 & -14.4945536 & 0.320014 & 0.000854 & 3 & Red Galaxy & 15.672 & 0.058 \\
		WISEPC J204101.64-142916.7 & 310.2568359 & -14.4879913 & 0.268144 & 1.17e-09 & 4 & Blue Galaxy & 14.717 & 0.039 \\
		WISEPC J204104.97-142724.2 & 310.2707214 & -14.4567318 & 0.313732 & 4.36e-05 & 4 & Green Galaxy & 15.405 & 0.051 \\
		WISEPC J204056.07-142557.8 & 310.2336426 & -14.4327354 & 0.396524 & 0.001 & 4 & VVDS S0 & 15.783 & 0.069 \\
		WISEPC J204052.21-142510.4 & 310.2175598 & -14.4195566 & 0.509227 & 0.00156 & 4 & VVDS Elliptical & 14.999 & 0.042 \\
		WISEPC J204051.15-142457.6 & 310.2131348 & -14.4160128 & 0.396874 & 8.68e-05 & 4 & VVDS Early Spiral & 15.699 & 0.062
	\enddata
	\tablenotetext{a}{The {\it WISE} magnitude used for selection.}
	\tablecomments{Example lines from the catalog of redshifts we will be making available based on data gathered with DEIMOS. Note that the selection magnitudes came from the L3o database used for internal verification and are thus extremely preliminary. All sources in the catalog have high quality ($Q\ge3$) redshifts. }
	\label{T:zcat}
\end{deluxetable}

\subsection{The Spectroscopically Unclassifiable Population}
The breakdown of the spectral qualities by sample can be found in Table \ref{T:Quals}. Table \ref{T:Detections} presents classifications rates as a function of spectrum type and bandpass for sources with magnitude uncertainties $\leq1$~mag. It is clear from the detection rates in \ref{T:Detections} that the most complete choice, overall, for analyzing the sample for which we could not get $Q>2$ spectra is to match $\mathrm{W}1$ with SDSS $r$, $i$, or $z$. Given the wavelength coverage of our spectra, we choose to use $r$ and $i$. 

\begin{figure} [t]
\begin{center}
	\includegraphics[width=\textwidth]{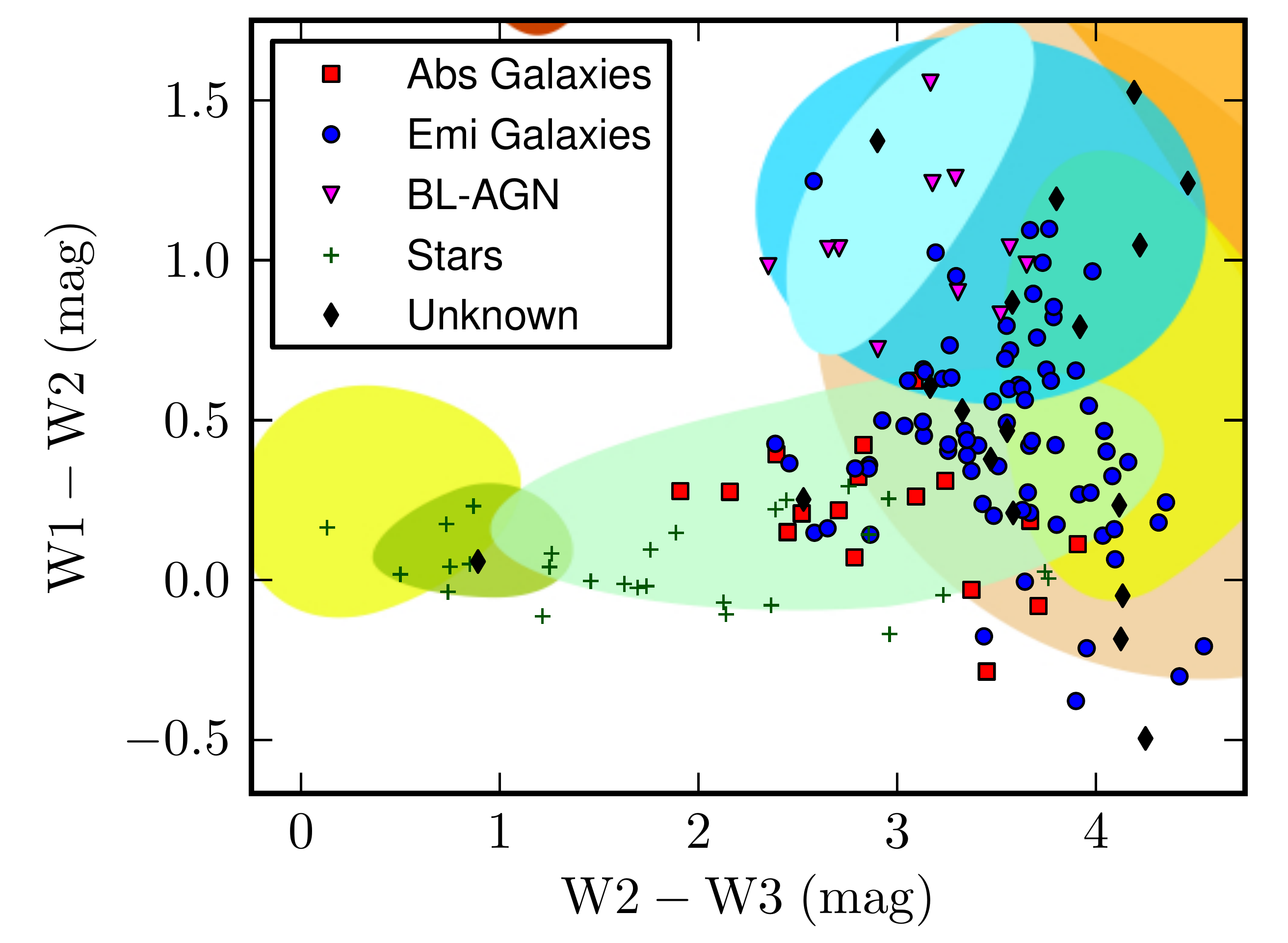}
	\caption{Our DEIMOS targets with minimally acceptable photometric uncertainty ($\sigma_{\mathrm{W}1},\ \sigma_{\mathrm{W}2},\ \mathrm{and}\ \sigma_{\mathrm{W}3} \leq 1.0$~mag). Sources marked with an
	 light grey circle (blue in the online version) are emission line galaxies with measurable redshifts, dark grey squares are absorption line galaxies (red in the online version), triangles pointing down (magenta in the online version) are broad-lined 
	 AGN, plus marks are stars (green in the online version), and black diamonds are sources we 
	 could not classify to a high degree of certainty based on the spectra we obtained.  Regions from Figure \ref{F:Bubble} are reproduced, in color, in the online version. }
\label{F:W123}
\end{center}
\end{figure}

We also provide Figure \ref{F:W123} for comparison with Figure \ref{F:Bubble}. While there are several stars outside of the boundary outlined for main sequence stars in Figure \ref{F:W123}, and this could be considered indicative of the presence of debris disks \citep{Wolf:2003}, these sources should be approached with caution. All of them have a W3 SNR in the range $(1.7, 3.5)$ and the size of the color excess varies nearly monotonically with the decrease in the W2 SNR. In short, these detections are near the noise floor of W3 and therefore the criterion that $\sigma_{W3}\le1.0$ is really a lower limit on the uncertainty in the flux in the direction of zero. Such Gaussian error estimates do not include the asymmetry required by a more rigorous treatment of the statistics. We also cannot rule out chance alignments with background objects that are dominating the {\it WISE} colors coinciding with a star that dominated the DEIMOS spectra.

\begin{figure*}[t]
\begin{center}
	\includegraphics[width=\textwidth]{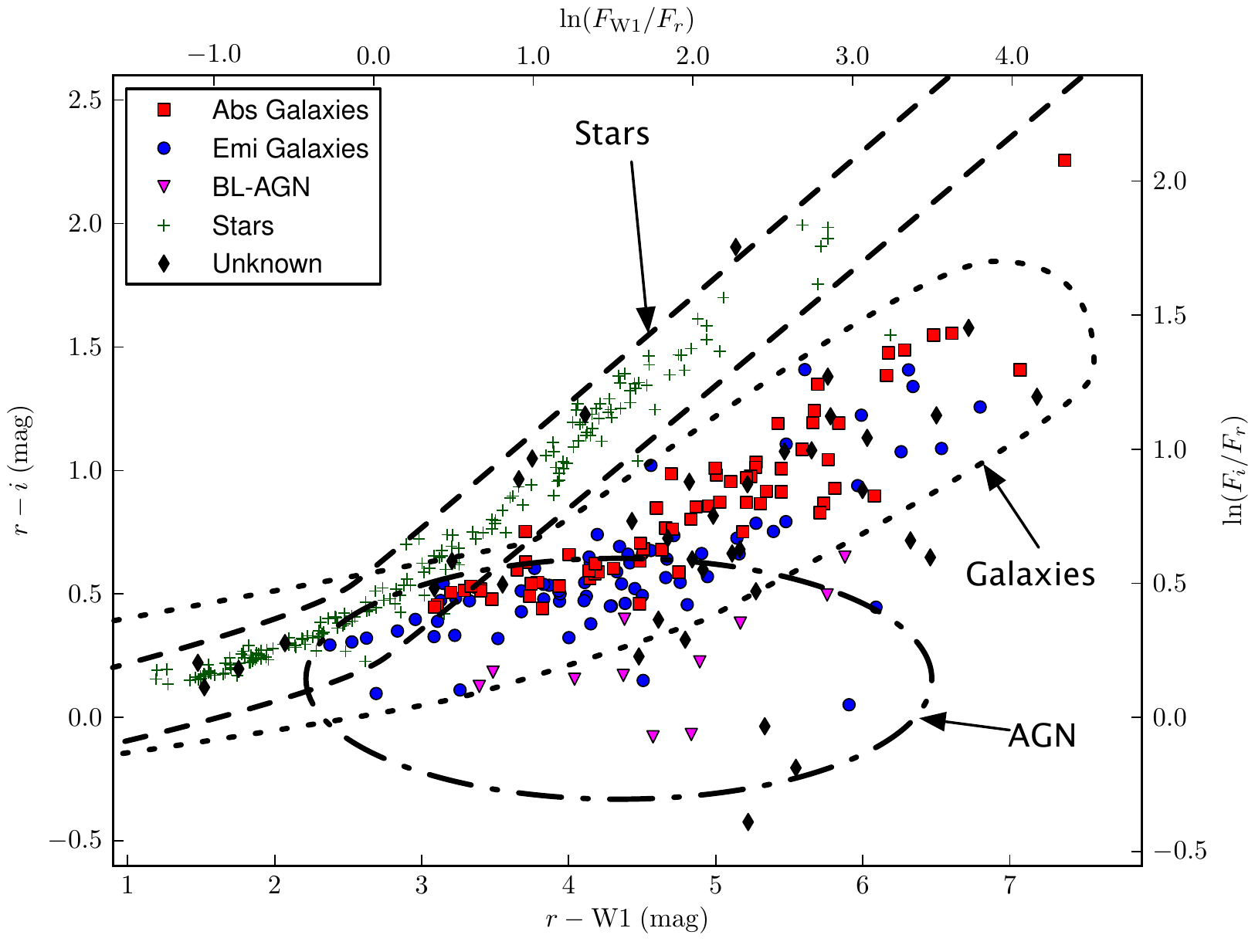}
	\caption{Our DEIMOS targets associated with a single target in SDSS DR8 
	and having minimally acceptable photometric uncertainty ($\sigma_i,\ \sigma_r,\ \mathrm{and}\ \sigma_{\mathrm{W}1} \leq 1.0$). Sources marked with an
	 open circle are galaxies with measurable redshifts, triangles pointing down are broad-lined 
	 AGN, plus marks are stars, and diamondss are sources we 
	 could not classify to a high degree of certainty based on the spectra we obtained.  We placed the regions qualitatively based on a combination of the data from this survey and a pseudo-random selection of spectroscopically classified sources from SDSS DR8 (not shown). }
\label{F:SColors}
\end{center}
\end{figure*}

Of the 762 not yet identified as spurious targets in our survey, 463 (61\%) are in the coverage of SDSS DR8. Of those covered, only 55 (12\%) did not have a clear counterpart in the SDSS DR8 database. From those 55 without a clear counterpart, we were able to get 20 (36\%) good spectra. Of the 408 targets with SDSS counterparts we failed to get good spectra for 29 (7\%). This has also given us the opportunity to characterize the sources that were in Sloan and not characterized by our survey, as shown in the color-color plot in Figure \ref{F:SColors}. There is an obvious bias toward missing targets that are in the predominantly extragalactic region of the color-color plot. This is consistent with the fact that the missed sources were overwhelmingly from the faintest sample, $\{<80\}$, and the findings in \cite{Jarrett:2011} that galaxies outnumber stars at the North Ecliptic Pole ($b\sim30^\circ$) for $\mathrm{W1} \gtrsim 15.0$~mag (W1 flux $\lesssim 300 \microJy$). Also of note is that the undetected galaxies are almost entirely from the region with $r-\mathrm{W}1 \ge 4.2$~mag, reinforcing the probability that our survey has a bias against high redshift sources.

We have listed in Table \ref{T:unconfirmed} all of the targets which had an $\mathrm{SNR} \ge 7.0$ in one of the {\it WISE} bands and for which there is no corresponding detection in the SDSS database, no Two Micron All Sky Survey (2MASS) detection, and no evidence of optical flux in the spectra we obtained. The purpose of Table \ref{T:unconfirmed} is to enable potential further followup of what are, at present, uniquely {\it WISE} sources.

\begin{deluxetable}{cccc}
	\tabletypesize{\scriptsize}
	\tablewidth{0.45\textwidth}
	\tablecaption{Spectroscopic Quality by Sample}
	\tablehead{\colhead{$Q$} & \colhead{$\{\ge120\}$} & \colhead{$\{80$--$120\}$} & 
		\colhead{$\{<80\}$} \\
		& \colhead{$(N_{\mathrm{tot}} = 449)$} & \colhead{$(N_{\mathrm{tot}} = 130)$} 
		& \colhead{$(N_{\mathrm{tot}} = 183)$}
	 }
	\startdata
		3--4 & 98.0\% & 84.6\% & 63.4\% \\
		0--2 & 2.0\% & 11.5\% & 18.6\% \\
		$-1$ & 0.0\% & 3.8\% & 18.0\% 
	\enddata
	\tablecomments{The fraction of targets in each spectral quality class broken 
	down into the W1 flux ranges that defined each sample, in $\microJy$. }
	\label{T:Quals}
\end{deluxetable}

\begin{deluxetable}{ccccc}
	\tabletypesize{\scriptsize}
	\tablewidth{0.75\textwidth}
	\tablecaption{Spectroscopically Classified Target Detection Rate by Channel}
	\tablehead{ \colhead{Filter} & \colhead{Emission Galaxies} & \colhead{Absorption Galaxies} & \colhead{Broad-lined AGN} & \colhead{Stars} \\
		\colhead{Name} & \colhead{$(f_{Q>2}\ [N_\mathrm{tot}])$} & \colhead{$(f_{Q>2}\ [N_\mathrm{tot}])$} & \colhead{$(f_{Q>2}\ [N_\mathrm{tot}])$} & \colhead{$(f_{Q>2}\ [N_\mathrm{tot}])$} 
	 }
	\startdata
		u & $58.8\%\ [97]$ & $29.1\%\ [79]$ & $91.7\%\ [12]$ & $79.0\%\ [210]$ \\
		g & $77.3\%\ [97]$ & $89.9\%\ [79]$ & $100.0\%\ [12]$ & $99.0\%\ [210]$ \\
		r & $80.4\%\ [97]$ & $91.1\%\ [79]$ & $100.0\%\ [12]$ & $99.0\%\ [210]$ \\
		i & $81.4\%\ [97]$ & $96.2\%\ [79]$ & $100.0\%\ [12]$ & $99.0\%\ [210]$ \\
		z & $80.4\%\ [97]$ & $94.9\%\ [79]$ & $100.0\%\ [12]$ & $99.0\%\ [210]$ \\
		J & $6.1\%\ [165]$ & $16.1\%\ [143]$ & $17.6\%\ [17]$ & $79.9\%\ [339]$ \\
		H & $6.1\%\ [165]$ & $16.1\%\ [143]$ & $11.8\%\ [17]$ & $78.5\%\ [339]$ \\
		K & $6.1\%\ [165]$ & $15.4\%\ [143]$ & $17.6\%\ [17]$ & $65.8\%\ [339]$ \\
		W1 & $98.2\%\ [165]$ & $100.0\%\ [143]$ & $100.0\%\ [17]$ & $100.0\%\ [339]$ \\
		W2 & $85.5\%\ [165]$ & $93.7\%\ [143]$ & $94.1\%\ [17]$ & $94.7\%\ [339]$ \\
		W3 & $50.9\%\ [165]$ & $11.9\%\ [143]$ & $64.7\%\ [17]$ & $8.8\%\ [339]$ \\
		W4 & $17.6\%\ [165]$ & $9.8\%\ [143]$ & $23.5\%\ [17]$ & $6.2\%\ [339]$
	\enddata
	\tablecomments{The fraction of targets with high quality ($Q > 2$) spectra detected ($\sigma \le 1.0$~mag) by each 
	photometric band, with the number of total available targets in brackets next to the fraction in 
	percent. The photometry used to construct this table came from SDSS DR8, 2MASS, and 
	{\it WISE}.}
	\label{T:Detections}
\end{deluxetable}

\begin{deluxetable}{lrrcccccc}
	\tabletypesize{\scriptsize}
	\rotate
	\tablewidth{0.83\textheight}
	\tablecaption{Well Detected Targets Without External Confirmation}
	\tablehead{\colhead{Designation} &\colhead{$\alpha$} & \colhead{$\delta$} & \colhead{$\mathrm{W1}$} &  
	\colhead{$\mathrm{W2}$} & \colhead{$\mathrm{W3}$} & \colhead{$\mathrm{W4}$} & 
	 \colhead{SDSS}\\
	& \colhead{(\deg)} & \colhead{(\deg)} & \colhead{ (mag)} & 
	\colhead{ (mag)} & \colhead{ (mag)} & 
	\colhead{ (mag)} & \colhead{DR8}} 
	\startdata
		WISEPC J223522.72+160246.4 & 338.844696 & 16.0462303 & $16.447 \pm 0.092$ & $15.936 \pm 0.185$ & $(12.879)$ & $(9.715)$ & Y \\
		WISEPC J223520.69+160256.2 & 338.8362122 & 16.0489521 & $16.37 \pm 0.092$ & $15.992 \pm 0.205$ & $12.522 \pm 0.406$ & $(9.236)$ & Y \\
		WISEPC J230337.04+040652.5 & 345.9043579 & 4.1145945 & $16.866 \pm 0.14$ & $16.445 \pm 0.334$ & $(12.196)$ & $(8.658)$ & Y \\
		WISEPC J014002.89-121449.2 & 25.0120697 & -12.2470074 & $16.629 \pm 0.126$ & $16.395 \pm 0.36$ & $12.278 \pm 0.382$ & $(8.869)$ & N \\
		WISEPC J023332.24+233608.5 & 38.3843384 & 23.6023865 & $16.664 \pm 0.126$ & $16.237 \pm 0.302$ & $(12.586)$ & $(9.168)$ & Y \\
		WISEPC J023331.58+233632.1 & 38.3816109 & 23.6089439 & $16.862 \pm 0.147$ & $16.07 \pm 0.264$ & $12.151 \pm 0.354$ & $(8.777)$ & Y \\
		WISEPC J023330.90+233924.9 & 38.378788 & 23.6569328 & $16.571 \pm 0.119$ & $15.834 \pm 0.217$ & $(12.598)$ & $(9.262)$ & Y \\
		WISEPC J032345.02+043501.2 & 50.9376183 & 4.5836883 & $16.091 \pm 0.08$ & $15.195 \pm 0.116$ & $(12.383)$ & $(9.418)$ & Y \\
		WISEPC J032342.02+043503.6 & 50.9251213 & 4.5843415 & $16.374 \pm 0.101$ & $15.427 \pm 0.148$ & $(12.605)$ & $(8.989)$ & Y \\
		WISEPC J032321.13+043126.1 & 50.8380814 & 4.5239253 & $16.897 \pm 0.155$ & $16.029 \pm 0.246$ & $12.45 \pm 0.452$ & $(8.883)$ & Y
	\enddata
	\tablecomments{The positions and {\it WISE} photometry for sources that lack confirming 
	detections in SDSS DR8, 2MASS, or this survey. All sources on this list must have 
	$\mathrm{SNR} \ge 7.0$ in at least one {\it WISE} channel and not be visually spurious to be 
	included.  Magnitudes in parentheses are $2$-$\sigma$ upper limits on the flux. All 
	photometry comes from the extremely preliminary reductions done to create the L3o 
	database. }
	\label{T:unconfirmed}
\end{deluxetable}

\section{Conclusion}
With its ability to detect $L_\star$ galaxies out to $z \sim 0.7$ in the areas with fewest repeat observations and its nearly $4\pi\ \mathrm{sr}$ coverage, {\it WISE} is able to play an important role in studies of galaxy populations and cosmic structures at moderate redshifts. The follow-on to this paper, \cite{Lake:2011-2}, reports on the $3.4\micron$ luminosity function to redshift $z\sim0.7$, derived using this data in conjunction with public spectroscopic data sets. In that paper we trace the contribution to extragalactic background light in the mid-IR from galaxies and constrain the stellar mass of galaxies. 

\acknowledgments{
This publication makes use of data products from the Wide-field Infrared Survey Explorer, which is a joint project of the University of California, Los Angeles, and the Jet Propulsion Laboratory/California Institute of Technology, funded by the National Aeronautics and Space Administration. The {\it WISE} website is \url{http://wise.ssl.berkeley.edu/}.

The analysis pipeline used to reduce the DEIMOS data was developed at UC Berkeley with support from NSF grant AST-0071048.

Funding for SDSS-III has been provided by the Alfred P. Sloan Foundation, the Participating Institutions, the National Science Foundation, and the U.S. Department of Energy. The SDSS-III web site is \url{http://www.sdss3.org/}.

SDSS-III is managed by the Astrophysical Research Consortium for the Participating Institutions of the SDSS-III Collaboration including the University of Arizona, the Brazilian Participation Group, Brookhaven National Laboratory, University of Cambridge, University of Florida, the French Participation Group, the German Participation Group, the Instituto de Astrofisica de Canarias, the Michigan State/Notre Dame/JINA Participation Group, Johns Hopkins University, Lawrence Berkeley National Laboratory, Max Planck Institute for Astrophysics, New Mexico State University, New York University, Ohio State University, Pennsylvania State University, University of Portsmouth, Princeton University, the Spanish Participation Group, University of Tokyo, University of Utah, Vanderbilt University, University of Virginia, University of Washington, and Yale University.

R.J.A. was supported by an appointment to the NASA Postdoctoral
Program at the Jet Propulsion Laboratory, administered by Oak Ridge
Associated Universities through a contract with NASA
}

\bibliography{ExpBib}

\begin{thebibliography}{27}
\expandafter\ifx\csname natexlab\endcsname\relax\def\natexlab#1{#1}\fi

\bibitem[{{Aihara} {et~al.}(2011){Aihara}, {Allende Prieto}, {An}, {Anderson},
  {Aubourg}, {Balbinot}, {Beers}, {Berlind}, {Bickerton}, {Bizyaev}, {Blanton},
  {Bochanski}, {Bolton}, {Bovy}, {Brandt}, {Brinkmann}, {Brown}, {Brownstein},
  {Busca}, {Campbell}, {Carr}, {Chen}, {Chiappini}, {Comparat}, {Connolly},
  {Cortes}, {Croft}, {Cuesta}, {da Costa}, {Davenport}, {Dawson}, {Dhital},
  {Ealet}, {Ebelke}, {Edmondson}, {Eisenstein}, {Escoffier}, {Esposito},
  {Evans}, {Fan}, {Femen{\'{\i}}a Castell{\'a}}, {Font-Ribera}, {Frinchaboy},
  {Ge}, {Gillespie}, {Gilmore}, {Gonz{\'a}lez Hern{\'a}ndez}, {Gott}, {Gould},
  {Grebel}, {Gunn}, {Hamilton}, {Harding}, {Harris}, {Hawley}, {Hearty}, {Ho},
  {Hogg}, {Holtzman}, {Honscheid}, {Inada}, {Ivans}, {Jiang}, {Johnson},
  {Jordan}, {Jordan}, {Kazin}, {Kirkby}, {Klaene}, {Knapp}, {Kneib},
  {Kochanek}, {Koesterke}, {Kollmeier}, {Kron}, {Lampeitl}, {Lang}, {Le Goff},
  {Lee}, {Lin}, {Long}, {Loomis}, {Lucatello}, {Lundgren}, {Lupton}, {Ma},
  {MacDonald}, {Mahadevan}, {Maia}, {Makler}, {Malanushenko}, {Malanushenko},
  {Mandelbaum}, {Maraston}, {Margala}, {Masters}, {McBride}, {McGehee},
  {McGreer}, {M{\'e}nard}, {Miralda-Escud{\'e}}, {Morrison}, {Mullally},
  {Muna}, {Munn}, {Murayama}, {Myers}, {Naugle}, {Fausti Neto}, {Cuong Nguyen},
  {Nichol}, {O'Connell}, {Ogando}, {Olmstead}, {Oravetz}, {Padmanabhan},
  {Palanque-Delabrouille}, {Pan}, {Pandey}, {P{\^a}ris}, {Percival},
  {Petitjean}, {Pfaffenberger}, {Pforr}, {Phleps}, {Pichon}, {Pieri}, {Prada},
  {Price-Whelan}, {Raddick}, {Ramos}, {Reyl{\'e}}, {Rich}, {Richards}, {Rix},
  {Robin}, {Rocha-Pinto}, {Rockosi}, {Roe}, {Rollinde}, {Ross}, {Ross},
  {Rossetto}, {S{\'a}nchez}, {Sayres}, {Schlegel}, {Schlesinger}, {Schmidt},
  {Schneider}, {Sheldon}, {Shu}, {Simmerer}, {Simmons}, {Sivarani}, {Snedden},
  {Sobeck}, {Steinmetz}, {Strauss}, {Szalay}, {Tanaka}, {Thakar}, {Thomas},
  {Tinker}, {Tofflemire}, {Tojeiro}, {Tremonti}, {Vandenberg}, {Vargas
  Maga{\~n}a}, {Verde}, {Vogt}, {Wake}, {Wang}, {Weaver}, {Weinberg}, {White},
  {White}, {Yanny}, {Yasuda}, {Yeche}, \& {Zehavi}}]{SDSSdr8}
{Aihara}, H., {et~al.} 2011, \apjs, 193, 29

\bibitem[{{Assef} {et~al.}(2010){Assef}, {Kochanek}, {Brodwin}, {Cool},
  {Forman}, {Gonzalez}, {Hickox}, {Jones}, {Le Floc'h}, {Moustakas}, {Murray},
  \& {Stern}}]{Assef:2010}
{Assef}, R.~J., {et~al.} 2010, \apj, 713, 970

\bibitem[{{Baldry} {et~al.}(2010){Baldry}, {Robotham}, {Hill}, {Driver},
  {Liske}, {Norberg}, {Bamford}, {Hopkins}, {Loveday}, {Peacock}, {Cameron},
  {Croom}, {Cross}, {Doyle}, {Dye}, {Frenk}, {Jones}, {van Kampen}, {Kelvin},
  {Nichol}, {Parkinson}, {Popescu}, {Prescott}, {Sharp}, {Sutherland},
  {Thomas}, \& {Tuffs}}]{Baldry:2010}
{Baldry}, I.~K., {et~al.} 2010, \mnras, 404, 86

\bibitem[{{Benford} {et~al.}(2011)}]{Benford:2011}
{Benford}, D., {et~al.} 2011, in prep.

\bibitem[{Bridge {et~al.}(2011)Bridge, Petty, Stern, Lake, Benford, Eisenhardt,
  Jarrett, Tsai, Wright, \& Blain}]{Bridge:2011}
Bridge, C., {et~al.} 2011, in prep.

\bibitem[{{Bruzual} \& {Charlot}(2003)}]{BC:2003}
{Bruzual}, G., \& {Charlot}, S. 2003, \mnras, 344, 1000

\bibitem[{{Colless} {et~al.}(2003){Colless}, {Peterson}, {Jackson}, {Peacock},
  {Cole}, {Norberg}, {Baldry}, {Baugh}, {Bland-Hawthorn}, {Bridges}, {Cannon},
  {Collins}, {Couch}, {Cross}, {Dalton}, {De Propris}, {Driver}, {Efstathiou},
  {Ellis}, {Frenk}, {Glazebrook}, {Lahav}, {Lewis}, {Lumsden}, {Maddox},
  {Madgwick}, {Sutherland}, \& {Taylor}}]{Colless:2003}
{Colless}, M., {et~al.} 2003, arXiv:astro-ph/0306581v1

\bibitem[{{Dai} {et~al.}(2009){Dai}, {Assef}, {Kochanek}, {Brodwin}, {Brown},
  {Caldwell}, {Cool}, {Dey}, {Eisenhardt}, {Eisenstein}, {Gonzalez}, {Jannuzi},
  {Jones}, {Murray}, \& {Stern}}]{Dai:2009}
{Dai}, X., {et~al.} 2009, \apj, 697, 506

\bibitem[{{Davis} {et~al.}(2003){Davis}, {Faber}, {Newman}, {Phillips},
  {Ellis}, {Steidel}, {Conselice}, {Coil}, {Finkbeiner}, {Koo}, {Guhathakurta},
  {Weiner}, {Schiavon}, {Willmer}, {Kaiser}, {Luppino}, {Wirth}, {Connolly},
  {Eisenhardt}, {Cooper}, \& {Gerke}}]{Davis:2003}
{Davis}, M., {et~al.} 2003, in SPIE Conference Series, ed. {P.~Guhathakurta},
  Vol. 4834, 161--172

\bibitem[{{Drinkwater} {et~al.}(2010){Drinkwater}, {Jurek}, {Blake}, {Woods},
  {Pimbblet}, {Glazebrook}, {Sharp}, {Pracy}, {Brough}, {Colless}, {Couch},
  {Croom}, {Davis}, {Forbes}, {Forster}, {Gilbank}, {Gladders}, {Jelliffe},
  {Jones}, {Li}, {Madore}, {Martin}, {Poole}, {Small}, {Wisnioski}, {Wyder}, \&
  {Yee}}]{Drinkwater:2010}
{Drinkwater}, M.~J., {et~al.} 2010, \mnras, 401, 1429

\bibitem[{{Eisenhardt} {et~al.}(2011){Eisenhardt}, {Assef}, {Benford}, {Blain},
  {Bridge}, J., {Cutri}, {Gelino}, {Griffith}, {Grillmair}, {Jarrett},
  {Lonsdale}, {Mason}, {McMillan}, {Petty}, {Sayers}, {Stanford}, {Stern},
  {Tsai}, {Wright}, {Wu}, \& {Yan}}]{Eisenhardt:2011}
{Eisenhardt}, P. R.~M., {et~al.} 2011, in prep.

\bibitem[{Feldman \& Cousins(1998)}]{Feldman:1998}
Feldman, G.~J., \& Cousins, R.~D. 1998, Physical Review D, 57, 3873

\bibitem[{{Jarosik} {et~al.}(2011){Jarosik}, {Bennett}, {Dunkley}, {Gold},
  {Greason}, {Halpern}, {Hill}, {Hinshaw}, {Kogut}, {Komatsu}, {Larson},
  {Limon}, {Meyer}, {Nolta}, {Odegard}, {Page}, {Smith}, {Spergel}, {Tucker},
  {Weiland}, {Wollack}, \& {Wright}}]{WMAP7}
{Jarosik}, N., {et~al.} 2011, \apjs, 192, 14

\bibitem[{{Jarrett} {et~al.}(1994){Jarrett}, {Dickman}, \&
  {Herbst}}]{Jarrett:1994}
{Jarrett}, T.~H., {Dickman}, R.~L., \& {Herbst}, W. 1994, \apj, 424, 852

\bibitem[{{Jarrett} {et~al.}(2011){Jarrett}, {Cohen}, {Masci}, {Wright},
  {Stern}, {Benford}, {Blain}, {Carey}, {Cutri}, {Eisenhardt}, {Lonsdale},
  {Mainzer}, {Marsh}, {Padgett}, {Petty}, {Ressler}, {Skrutskie}, {Stanford},
  {Surace}, {Tsai}, {Wheelock}, \& {Yan}}]{Jarrett:2011}
{Jarrett}, T.~H., {et~al.} 2011, \apj, 735, 112

\bibitem[{{Jones} {et~al.}(2009){Jones}, {Read}, {Saunders}, {Colless},
  {Jarrett}, {Parker}, {Fairall}, {Mauch}, {Sadler}, {Watson}, {Burton},
  {Campbell}, {Cass}, {Croom}, {Dawe}, {Fiegert}, {Frankcombe}, {Hartley},
  {Huchra}, {James}, {Kirby}, {Lahav}, {Lucey}, {Mamon}, {Moore}, {Peterson},
  {Prior}, {Proust}, {Russell}, {Safouris}, {Wakamatsu}, {Westra}, \&
  {Williams}}]{Jones:2009}
{Jones}, D.~H., {et~al.} 2009, \mnras, 399, 683

\bibitem[{{Kochanek} {et~al.}(2011)}]{AGES}
{Kochanek}, C.~S., {et~al.} 2011, in prep.

\bibitem[{{Lake} {et~al.}(2011)}]{Lake:2011-2}
{Lake}, S.~E., {et~al.} 2011, in prep.

\bibitem[{{Liu} {et~al.}(2008){Liu}, {Cutri}, {Greanias}, {Duval},
  {Eisenhardt}, {Elwell}, {Heinrichsen}, {Howard}, {Irace}, {Mainzer},
  {Razzaghi}, {Royer}, \& {Wright}}]{Liu:2008}
{Liu}, F., {et~al.} 2008, in SPIE Conference Series, Vol. 7017

\bibitem[{{Masters} \& {Capak}(2011)}]{Masters:2011}
{Masters}, D., \& {Capak}, P. 2011, \pasp, 123, 638

\bibitem[{{Oke} {et~al.}(1995){Oke}, {Cohen}, {Carr}, {Cromer}, {Dingizian},
  {Harris}, {Labrecque}, {Lucinio}, {Schaal}, {Epps}, \& {Miller}}]{LRIS}
{Oke}, J.~B., {et~al.} 1995, \pasp, 107, 375

\bibitem[{{Phillips} {et~al.}(2002)}]{DEIMOS}
{Phillips}, A.~C., {et~al.} 2002, in Bulletin of the American Astronomical
  Society, Vol.~34, AAS Meeting Abstracts, 137.02

\bibitem[{{Polletta} {et~al.}(2007){Polletta}, {Tajer}, {Maraschi},
  {Trinchieri}, {Lonsdale}, {Chiappetti}, {Andreon}, {Pierre}, {Le F{\`e}vre},
  {Zamorani}, {Maccagni}, {Garcet}, {Surdej}, {Franceschini}, {Alloin},
  {Shupe}, {Surace}, {Fang}, {Rowan-Robinson}, {Smith}, \&
  {Tresse}}]{Polletta:2007}
{Polletta}, M., {et~al.} 2007, \apj, 663, 81

\bibitem[{{Scoville} {et~al.}(2007){Scoville}, {Aussel}, {Brusa}, {Capak},
  {Carollo}, {Elvis}, {Giavalisco}, {Guzzo}, {Hasinger}, {Impey}, {Kneib},
  {LeFevre}, {Lilly}, {Mobasher}, {Renzini}, {Rich}, {Sanders}, {Schinnerer},
  {Schminovich}, {Shopbell}, {Taniguchi}, \& {Tyson}}]{Lilly:2007}
{Scoville}, N., {et~al.} 2007, \apjs, 172, 1

\bibitem[{{Silva} {et~al.}(1998){Silva}, {Granato}, {Bressan}, \&
  {Danese}}]{Silva:1998}
{Silva}, L., {Granato}, G.~L., {Bressan}, A., \& {Danese}, L. 1998, \apj, 509,
  103

\bibitem[{{Wolf} \& {Hillenbrand}(2003)}]{Wolf:2003}
{Wolf}, S., \& {Hillenbrand}, L.~A. 2003, \apj, 596, 603

\bibitem[{{Wright} {et~al.}(2010){Wright}, {Eisenhardt}, {Mainzer}, {Ressler},
  {Cutri}, {Jarrett}, {Kirkpatrick}, {Padgett}, {McMillan}, {Skrutskie},
  {Stanford}, {Cohen}, {Walker}, {Mather}, {Leisawitz}, {Gautier}, {McLean},
  {Benford}, {Lonsdale}, {Blain}, {Mendez}, {Irace}, {Duval}, {Liu}, {Royer},
  {Heinrichsen}, {Howard}, {Shannon}, {Kendall}, {Walsh}, {Larsen}, {Cardon},
  {Schick}, {Schwalm}, {Abid}, {Fabinsky}, {Naes}, \& {Tsai}}]{Wright:2010}
{Wright}, E.~L., {et~al.} 2010, \aj, 140, 1868

\end{thebibliography}

\end{document}